\documentclass[english,aps,prb,twocolumn,longbibliography]{revtex4-2}
\usepackage[T1]{fontenc}
\usepackage[latin9]{inputenc}
\usepackage{color}
\usepackage{amsmath}
\usepackage{amssymb}
\usepackage{graphicx}
\usepackage{esint}

\makeatletter
\@ifundefined{textcolor}{}
{%
 \definecolor{BLACK}{gray}{0}
 \definecolor{WHITE}{gray}{1}
 \definecolor{RED}{rgb}{1,0,0}
 \definecolor{GREEN}{rgb}{0,1,0}
 \definecolor{BLUE}{rgb}{0,0,1}
 \definecolor{CYAN}{cmyk}{1,0,0,0}
 \definecolor{MAGENTA}{cmyk}{0,1,0,0}
 \definecolor{YELLOW}{cmyk}{0,0,1,0}
 }

\makeatother

\usepackage{babel}
\begin{document}

\newcommand{\1}{{\bf \scriptstyle 1}\!\!{1}}
\newcommand{\unit}{\overleftrightarrow{{\bf \scriptstyle 1}\!\!{1}}}
\newcommand{\I}{{\rm i}}
\newcommand{\p}{\partial}
\newcommand{\D}{^{\dagger}}
\newcommand{\hbe}{\hat{\bf e}}
\newcommand{\bfa}{{\bf a}}
\newcommand{\bx}{{\bf x}}
\newcommand{\hbx}{\hat{\bf x}}
\newcommand{\by}{{\bf y}}
\newcommand{\hby}{\hat{\bf y}}
\newcommand{\br}{{\bf r}}
\newcommand{\hbr}{\hat{\bf r}}
\newcommand{\bj}{{\bf j}}
\newcommand{\bk}{{\bf k}}
\newcommand{\bn}{{\bf n}}
\newcommand{\bv}{{\bf v}}
\newcommand{\bp}{{\bf p}}
\newcommand{\tp}{\tilde{p}}
\newcommand{\tbp}{\tilde{\bf p}}
\newcommand{\bu}{{\bf u}}
\newcommand{\hbz}{\hat{\bf z}}
\newcommand{\bA}{{\bf A}}
\newcommand{\calA}{\mathcal{A}}
\newcommand{\calB}{\mathcal{B}}
\newcommand{\tC}{\tilde{C}}
\newcommand{\bD}{{\bf D}}
\newcommand{\bE}{{\bf E}}
\newcommand{\calF}{\mathcal{F}}
\newcommand{\bB}{{\bf B}}
\newcommand{\bG}{{\bf G}}
\newcommand{\calG}{\mathcal{G}}
\newcommand{\obG}{\overleftrightarrow{\bf G}}
\newcommand{\bJ}{{\bf J}}
\newcommand{\bK}{{\bf K}}
\newcommand{\bL}{{\bf L}}
\newcommand{\tL}{\tilde{L}}
\newcommand{\bP}{{\bf P}}
\newcommand{\calP}{\mathcal{P}}
\newcommand{\calR}{\mathcal{R}}
\newcommand{\bQ}{{\bf Q}}
\newcommand{\bR}{{\bf R}}
\newcommand{\bS}{{\bf S}}
\newcommand{\bH}{{\bf H}}
\newcommand{\balpha}{\mbox{\boldmath $\alpha$}}
\newcommand{\talpha}{\tilde{\alpha}}
\newcommand{\bsigma}{\mbox{\boldmath $\sigma$}}
\newcommand{\hbeta}{\hat{\mbox{\boldmath $\eta$}}}
\newcommand{\bSigma}{\mbox{\boldmath $\Sigma$}}
\newcommand{\bomega}{\mbox{\boldmath $\omega$}}
\newcommand{\bpi}{\mbox{\boldmath $\pi$}}
\newcommand{\bphi}{\mbox{\boldmath $\phi$}}
\newcommand{\hbphi}{\hat{\mbox{\boldmath $\phi$}}}
\newcommand{\btheta}{\mbox{\boldmath $\theta$}}
\newcommand{\hbtheta}{\hat{\mbox{\boldmath $\theta$}}}
\newcommand{\hbxi}{\hat{\mbox{\boldmath $\xi$}}}
\newcommand{\hbzeta}{\hat{\mbox{\boldmath $\zeta$}}}
\newcommand{\brho}{\mbox{\boldmath $\rho$}}
\newcommand{\bnabla}{\mbox{\boldmath $\nabla$}}
\newcommand{\bmu}{\mbox{\boldmath $\mu$}}
\newcommand{\bepsilon}{\mbox{\boldmath $\epsilon$}}

\newcommand{\iLambda}{{\it \Lambda}}
\newcommand{\cL}{{\cal L}}
\newcommand{\cH}{{\cal H}}
\newcommand{\cU}{{\cal U}}
\newcommand{\cT}{{\cal T}}

\newcommand{\be}{\begin{equation}}
\newcommand{\ee}{\end{equation}}
\newcommand{\bea}{\begin{eqnarray}}
\newcommand{\eea}{\end{eqnarray}}
\newcommand{\beqa}{\begin{eqnarray*}}
\newcommand{\eeqa}{\end{eqnarray*}}
\newcommand{\nn}{\nonumber}
\newcommand{\DD}{\displaystyle}

\newcommand{\ba}{\begin{array}{c}}
\newcommand{\baa}{\begin{array}{cc}}
\newcommand{\baaa}{\begin{array}{ccc}}
\newcommand{\baaaa}{\begin{array}{cccc}}
\newcommand{\ea}{\end{array}}

\newcommand{\bma}{\left[\begin{array}{c}}
\newcommand{\bmaa}{\left[\begin{array}{cc}}
\newcommand{\bmaaa}{\left[\begin{array}{ccc}}
\newcommand{\bmaaaa}{\left[\begin{array}{cccc}}
\newcommand{\ema}{\end{array}\right]}

\title{\textcolor{black}{Plasmonically enhanced tunable spectrally selective NIR and SWIR photodetector based on intercalation doped nanopatterned multilayer graphene}}

\author{\textcolor{black}{Muhammad Waqas Shabbir$^{(1)}$}}

\author{\textcolor{black}{Michael N. Leuenberger$^{(1,2)}$}}

\email{michael.leuenberger@ucf.edu}

\affiliation{$^{(1)}$ NanoScience Technology Center and Department of Physics, University of Central Florida, Orlando, FL 32826, USA. \\
$^{(2)}$ College of Optics and Photonics, University of Central Florida, Orlando, FL 32826, USA.}

\begin{abstract}
We present a proof of concept for a spectrally selective near-infrared (NIR) and short-wavelength infrared (SWIR) photodetector based on nanopatterned multilayer graphene intercalated with FeCl$_3$ (NPMLG-FeCl$_3$), enabling large modulation p-doping of graphene.
The localized surface plasmons (LSPs) on the graphene sheets in NPMLG-FeCl$_3$ allow for electrostatic tuning of the photodetection in the NIR and SWIR regimes from $\lambda =1.3$ $\mu$m to 3 $\mu$m, which is out of range for nanopatterned monolayer graphene (NPG).
Most importantly, the LSPs along with an optical cavity increase the absorbance from about $N\times 2.6$\% for $N$-layer graphene-FeCl$_3$ (without patterning) to nearly 100\% for NPMLG-FeCl$_3$, where the strong absorbance occurs locally inside the graphene sheets only.
Our NIR and SWIR detection scheme relies on the photo-thermoelectric effect induced by asymmetric patterning of the multi-layer graphene (MLG) sheets. The LSPs on the nanopatterned side create hot carriers that give rise to Seebeck photodetection at room temperature achieving a responsivity of $\calR=6.15\times 10^3$ V/W, a detectivity of $D^*=2.3\times 10^{9}$ Jones, and an ultrafast response time of the order of 100 ns.
Our theoretical results pave the way to graphene-based photodetection, optical IR communication,  IR color displays, and IR spectroscopy in the NIR, SWIR, mid-wavelength infrared (MWIR), and long-wavelength infrared (LWIR) regimes.

\textcolor{black}{KEYWORDS: Localized surface plasmons, graphene, light absorption, Seebeck effect, infrared light detection. }
\end{abstract}
\maketitle

Because of the weak absorbance of pristine graphene of around 2\%, we created nanopatterned CVD-grown single-layer graphene (NPG) that exhibits absorbance exceeding 60\% in the long-wavelength infrared (LWIR) regime between $\lambda = 8$ $\mu$m to 12 $\mu$m.\cite{Safaei2017,SafaeiACS} Recently, we showed that NPG with smaller sizes of hexagonally arranged holes and smaller lattice constants exhibits absorbance of 80\% in the mid-wavelength (MWIR) regime between $\lambda =3$ $\mu$m and 8 $\mu$m.
Due to the resolution limit of e-beam lithogaphic systems it is currently impossible to create smaller nanopatterns for increasing the absorbance of graphene using nanopatterning at shorter wavelengths.

Here, we show that  LSP resonances can be realized in the technologically relevant NIR and SWIR regimes between $\lambda =1.3$ $\mu$m and 3 $\mu$m by means of nanopatterned multilayer graphene (NPMLG)  intercalated with FeCl$_3$. Multilayer graphene intercalated with ferric chloride FeCl$_3$, dubbed graphexeter, was created for realizing an all-graphene photodetector which operates at around 6 $\mu$m.\cite{Withers2013} The intriguing proximity effect of single layers of FeCl$_3$ between the graphene sheets is to p-dope the graphene sheets to a Fermi energy of $E_F=-0.6$ eV due to the large work function of FeCl$_3$ of $W_{{\rm FeCl}_3}=5.1$ eV and the resulting charge transfer between FeCl$_3$ and graphene, which has a work function of around $W_g=4.6$ eV.\cite{Bointon2015,Jiang2017}  
Remarkably, the bandstructures of graphene and FeCl$_3$ remain completely decoupled due to the incommensurate lattice structures, i.e. their lattice constants are 2.46 \AA and 6.06 \AA, respectively.\cite{Zhan2010} 
According to Ref.~\onlinecite{Zhan2010}, MLG-FeCl$_3$ has a gap of 1.2 eV, to which the Fermi energy can be tuned by n-doping, corresponding to a wavelength of 1.03 $\mu$m,  and is therefore transparent in the NIR and SWIR regimes between $\lambda =1.3$ $\mu$m and 3 $\mu$m.
The refractive index of FeCl$_3$ is $n=1.365$, which gives a dielectric constant of $\varepsilon=n^2=1.86$.
By creating a hexagonal nanopattern of holes inside the multilayer graphene/FeCl$_3$-intercalated heterostructure (NPMLG-FeCl$_3$), we show that it is possible to achieve LSP resonances in the NIR and SWIR regimes between $\lambda =1.3$ $\mu$m to 3 $\mu$m with absorbances of nearly 100\%.
The method to tune the spectrally selective absorbance in NPMLG by means of a gate voltage $V_g$ is based on the fact that $V_g$ varies
the Fermi energy $E_F$ inside NPMLG, thereby varying the charge density and therefore resonance wavelength of the LSPs around the circular holes
in the wavelength regime between 1.3 $\mu$m and 3 $\mu$m.

\begin{figure*}[htb]
\begin{centering}
\includegraphics[width=18.0cm]{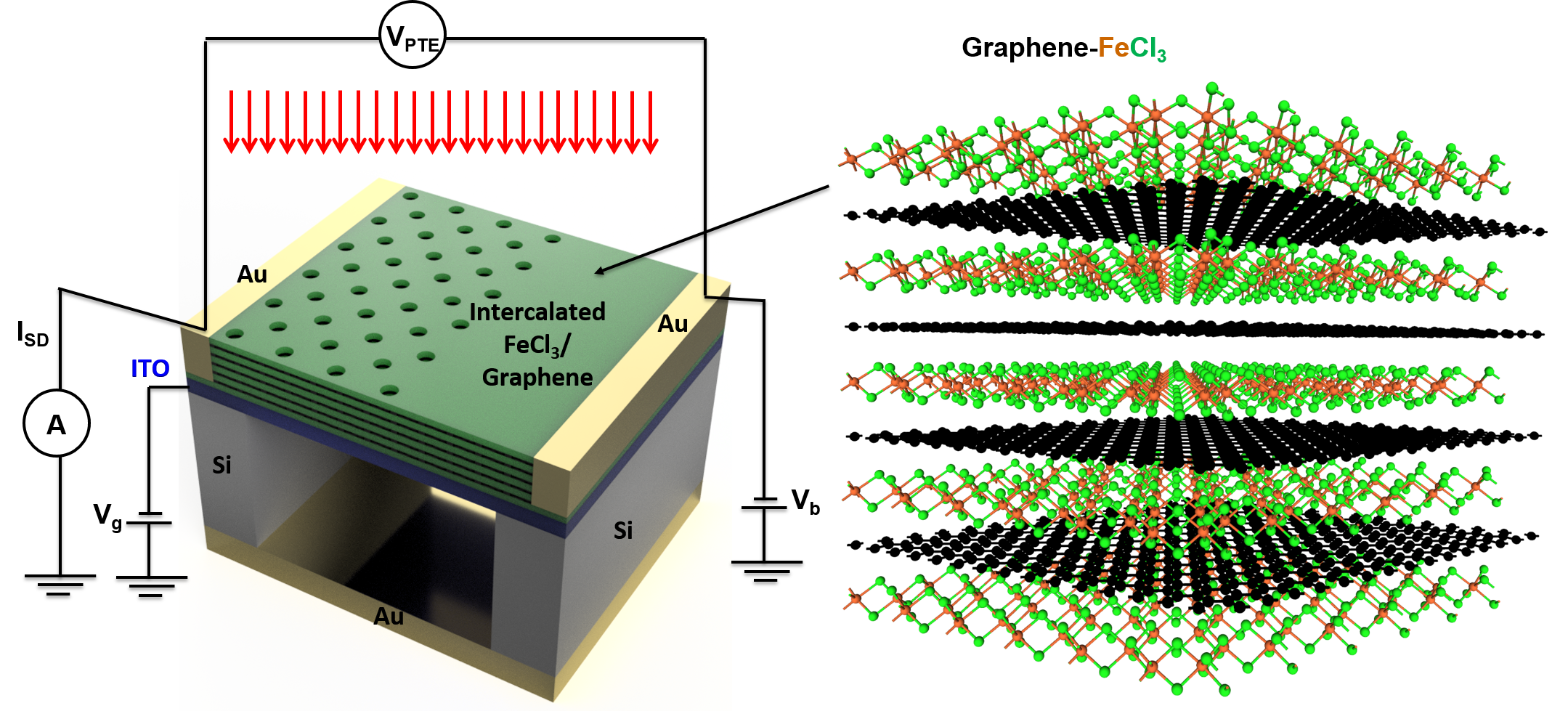}
\end{centering}
\caption{Schematic showing our proposed ultrafast NIR photodetector based on suspended NPG intercalated with FeCl$_3$ placed on top of a cavity, which can be tuned by means of a gate voltage applied to the ITO layer. Here we show the example using five graphene layers for MLG-FeCl$_3$ for stage 1 intercalation.  
\label{fig:NPMLG_photodetector} }
\end{figure*}

Taking advantage of the special properties of FeCl$_3$, we present the proof of concept for a NIR photodetector based on NPMLG intercalated with FeCl$_3$. 
Fig.~\ref{fig:NPMLG_photodetector} shows the schematic of our envisioned NPMLG-FeCl$_3$ photodetector. When the incident NIR or SWIR light field is maximized at the position of the NPMLG-FeCl$_3$ heterostructure by means of the Au mirror and also in resonance with the LSPs around the holes of the NPMLG, the NIR or SWIR light gets strongly absorbed with absorbance exceeding 95\%, as shown in Fig.~\ref{fig:NPMLG_absorbance}. Subsequently, the LSPs decay by creating hot carriers inside the graphene sheets due to boundary-assisted intraband Landau damping.\cite{Yan2013} 
While it is possible to detect the NIR or SWIR light by means of the bolometric effect relying on the change in conductance through NPG, this method is relatively slow because the lattice of NPG needs to be heated by means of the electron-phonon interaction before a signal can be detected, which is typically of the order of 1 ms.\cite{Shabbir2021_VO2}
In order to reduce the response time to about 100 ns, we choose to nanopattern only about half of the area of each graphene sheet, as shown in Fig.~\ref{fig:NPMLG_photodetector}.
Due to the asymmetric heating inside each partially nanopatterned graphene, a temperature gradient is created, which leads to the diffusion of the hot carriers from the nanopatterned side to the pristine side of the graphene sheet. This charge motion yields a Seebeck voltage across the Au source-drain contacts, giving rise to the plasmonically enhanced photothermoelectric effect inside each partically nanopatterned graphene sheet.
Due to the record-high temperature difference $\Delta T=5$ K, our envisioned NPMLG-FeCl$_3$ NIR and SWIR photodetectors exhibit extraordinarily large responsivity of $\calR=6.15\times 10^3$ V/W and detectivity of $D^*=2.33\times 10^{9}$ Jones.

As mentioned above, due to limited resolution of e-beam lithography it is currently impossible to realize LSP resonances in NPG in the NIR or SWIR.
A natural way to increase the plasmon frequency is to increase the charge density of the material.
In graphene this can be achieved by stacking several graphene sheets on top of each other.
While for small twist angles between $\theta=0^\circ$ up to about $15^\circ$, twisted bilayer, trilayer, and multilayer graphene exhibit interlayer coupling, resulting in e.g. electrostatically tunable band gaps in bilayer graphene\cite{Zhang2009} at $\theta=0$ and exotic many-body correlations, such as superconductivity, in twisted bilayer graphene at magic angle $\theta=1.1^\circ$,\cite{Cao2018} twist angles around $\theta=30^\circ$ in twisted bilayer graphene suppress completely interlayer coupling due to the mismatch of the k-space locations of the Dirac cones of the two layers.\cite{Wang2010}
At first, such incommensurate stackings of graphene layers might look like being ideal candidates for increasing the plasmon frequency of the LSPs.
However, the electrostatic doping of all the incommensurate graphene layers is impossible due to screening and also due to electronic decoupling of the layers.\cite{Hass2008}

\begin{figure*}[htb]
\begin{centering}
\includegraphics[width=\linewidth]{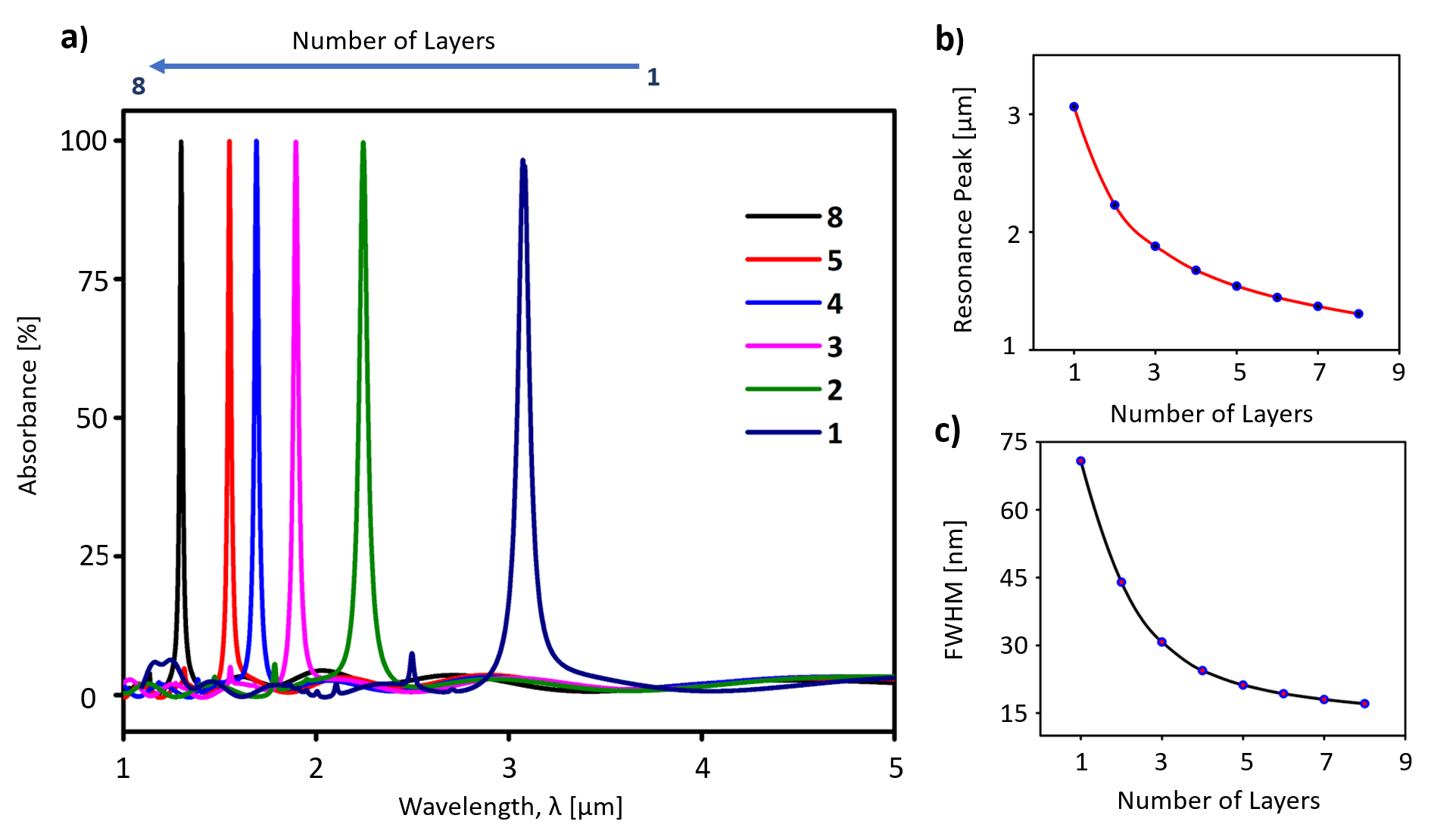}
\end{centering}
\caption{Absorbance $A(\lambda)$ of the NPMLG-FeCl$_3$ heterostructure shown in Figs.~\ref{fig:NPMLG_photodetector} with Fermi energy $E_F=-1.0$ eV, mobility $\mu=1500$ V/cm$^2$s, hole diameter of $a=40$ nm and period $\calP=60$ nm at $T=300$ K. The results are obtained by FDTD calculations. (a) For a single NPG sheet the LSP resonance is at around 4.5 $\mu$m. At $N=5$, the LSP resonance is at $1.55$ $\mu$m. At $N=8$, the LSP resonance is at $1.30$ $\mu$m. (b) The LSP resonance gets shorter and shorter with larger number $N$ of NPG layers. (c) The full width at half maximum (FWHM) of the LSP resonance gets smaller and smaller for larger $N$.
\label{fig:NPMLG_absorbance} }
\end{figure*}

In the seminal paper by Withers et al.\cite{Withers2013} the authors studied the optoelectronic properties of graphene/FeCl$_3$-intercalated few layer graphene, which they called graphexeter. They demonstrated a photodetector based on the photothermoelectric effect with a responsivitiy of $\calR=0.1$ V/W at a wavelength of around $\lambda=6$ $\mu$m.  Interest in such intercalation doped multilayer graphene heterostructures has been recently rekindled by the realization of modulation doping of multilayer graphene by means of $\alpha$-RuCl$_3$.\cite{Wang2020} Due to the large work function of $\alpha$-RuCl$_3$, $W_{\alpha-{\rm RuCl}_3}=6.1$ eV, substantial p-doping of all graphene layers of around $E_F=-0.8$ eV can be achieved without electrostatic gating.
There are conflicting reports on the band gap of RuCl$_3$. In Ref.~\onlinecite{Reschke2017} IR reflectivity and transmission measurements reveal a band gap of 200 meV, which agrees with the energies of spin-orbit excitons measured in Ref.~\onlinecite{Warzanowski2020}. However, photoemission and inverse photoemission spectroscopies find a much larger band gap of 1.9 eV.\cite{Sinn2016}

Since RuCl$_3$ has an optical band gap of the order of 200 meV,\cite{Reschke2017} it is transparent only up to a wavelength of $\lambda=3$ $\mu$m. 
In contrast, FeCl$_3$-intercalated few layer graphene MLG-FeCl$_3$ is transparent in the visible, infrared, and THz regimes due to the large band gap of FeCl$_3$.\cite{Khrapach2012,Zhukova2019}
The absorbance of MLG-FeCl$_3$ is approximately $N\times 2.6$\%, where $N$ is the number of layers.\cite{Khrapach2012}
This is the reason why MLG-FeCl$_3$ can be used as a transparent electrode in the visible, infrared, and THz regimes.
Taking advantage of the LSP resonances in nanopatterned graphene, we can increase the absorbance of NPMLG-FeCl$_3$ in the NIR and SWIR regimes to nearly 100\%, as shown in Fig.~\ref{fig:absorbance_EF}. Therefore, we propose the novel heterostructure material
NPMLG-FeCl$_3$ for creating a NIR and SWIR photodetectors based on the plasmonically enhanced photothermoelectric effect, which is shown in Fig.~\ref{fig:NPMLG_photodetector}.
The MLG channel length, which is the distance between the Au source and drain contacts, is chosen to be $L=10$ $\mu$m, of the same order as the diffusion length of charge carriers in graphene.\cite{Song2011}
The MLG channel width is chosen to be $W=10$ $\mu$m as well because carrier collection does not need to be enhanced.
The optimized hexagonal nanopattern on about half of the MLG-FeCl$_3$ heterostructure (see Fig.~\ref{fig:NPMLG_photodetector}) has a period of 60 nm and a hole diameter of 40 nm.

\begin{figure*}[htb]
\begin{centering}
\includegraphics[width=\linewidth]{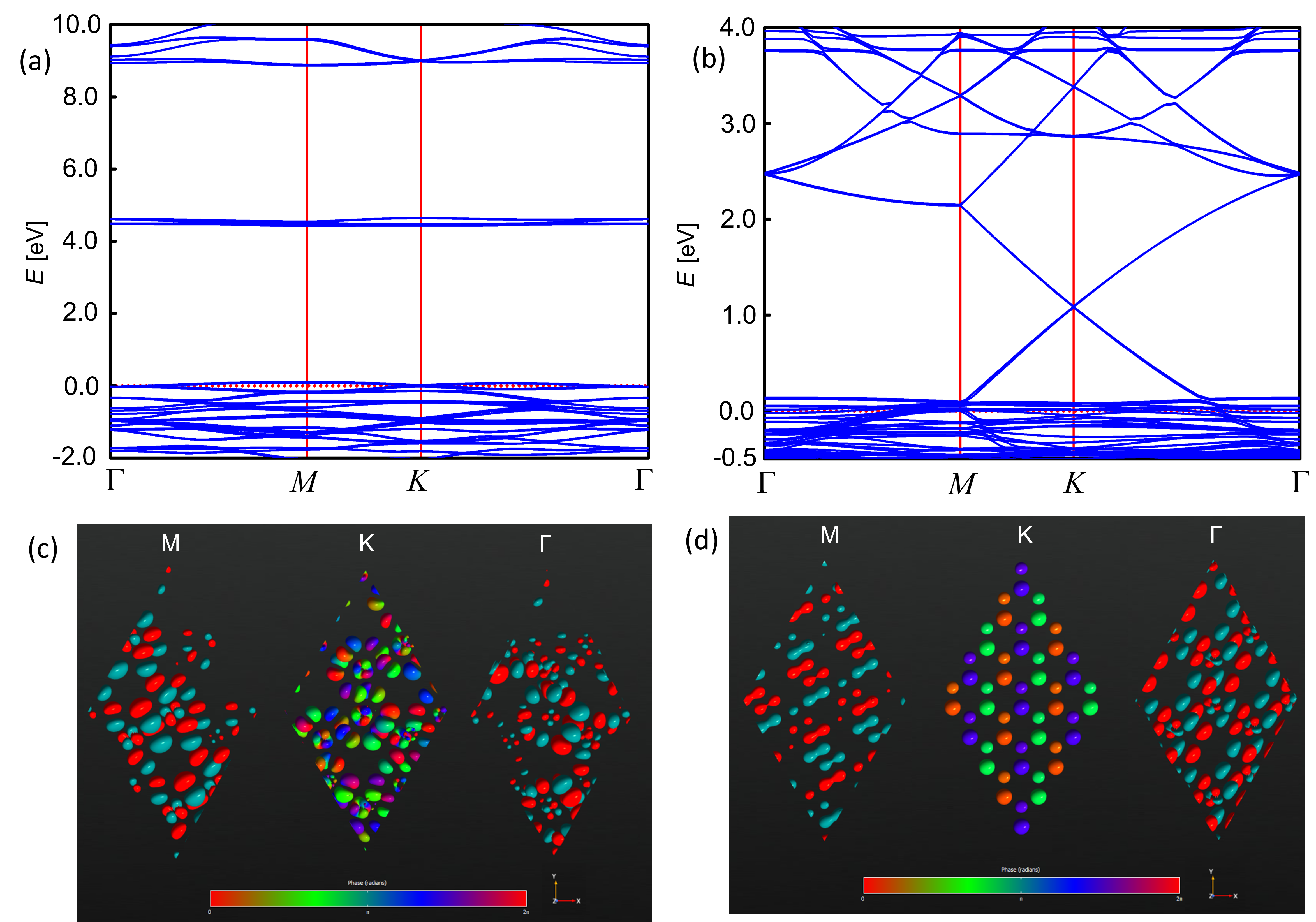}
\end{centering}
\caption{(a) Band structure of bulk FeCl$_3$. (b) Band structure of MLG-FeCl$_3$. (c) Bloch states localized in FeCl$_3$ in MLG-FeCl$_3$ at M, K, and $\Gamma$ points. (d) Bloch states localized in graphene in MLG-FeCl$_3$ at M, K, and $\Gamma$ points.
\label{fig:bandstructures} }
\end{figure*}

For the subsequent finite-difference time domain (FDTD) calculations, we perform first ab-initio density-functional theory (DFT) calculations to determine the band structure of MLG-FeCl$_3$.
The generalized gradient approximation (GGA) plus Hubbard-U parameter is used with the Perdew-Burke-Ernzerhof (PBE) parametrization\cite{PZ_functionals} of the correlation energy. 
The calculations are implemented within the Synopsis Atomistix Toolkit (ATK) 2019.12.\cite{QW_1}
As shown in Fig.~\ref{fig:NPMLG_photodetector}, we consider MLG-FeCl$_3$ with stage 1 intercalation, where single graphene layers are sandwiched by single FeCl$_3$ layers. 
We consider AB stacking between adjacent graphene layers, similar to pristine graphite, 
and ABC stacking between adjacent FeCl$_3$ layers, similar to pristine bulk FeCl$_3$.\cite{Hashimoto1989} 
The translational shifts between graphene layers and also between FeCl$_3$ layers ensures complete decoupling of the electron states between the graphene-graphene, graphene-FeCl$_3$, and FeCl$_3$-FeCl$_3$ layers.
Following Ref.~\onlinecite{Li2013}, we reduce the lattice mismatch by using a supercell with 2x2 periods of FeCl$_3$ and 5x5 periods of graphene.
The results of the DFT GGA+U band structure calculation are shown in Fig.~\ref{fig:bandstructures}, which are similar to the results obtained in Ref.~\onlinecite{Li2013}.
A choice of $U=6$ eV results in a FeCl$_3$ band gap of around 4.5 eV [see Fig.~\ref{fig:bandstructures}(a)], which agrees well with experiments.\cite{Khrapach2012,Zhukova2019}
The difference is that by incorporating a relative translational shift between each pair of adjacent layers, we achieve complete decoupling between the electron states of each layer.
This is clearly visible for Bloch states localized in FeCl$_3$ in MLG-FeCl$_3$ [see Fig.~\ref{fig:bandstructures}(c)] and Bloch states localized in graphene in MLG-FeCl$_3$ [see Fig.~\ref{fig:bandstructures}(d)].
The band structure of MLG-FeCl$_3$ shown in Fig.~\ref{fig:bandstructures}(b) shows clearly strong p-doping by shifting the Fermi energy to $E_F=-1.0$ eV relative to the graphene Dirac point.

In order to perform the FDTD, we use the linear dispersion relation of graphene, which gives rise to the intraband optical conductivity\cite{Safaei2017,Paudel2017}
\be
\sigma _{\rm intra}(\omega ) = \frac{e^2}{\pi\hbar^2}\frac{2k_BT}{\tau^{-1} - i\omega}\ln \left[ 2\cosh \left( \frac{\varepsilon _F}{2k_BT} \right) \right],
\ee
 which in the case of ${\varepsilon _F} \gg {k_B}T$   is reduced to 
 \be
\sigma_{\rm intra}(\omega) = \frac{e^2}{\pi\hbar^2}\frac{E_F}{\tau^{-1} - i\omega }=\frac{2\varepsilon_m\omega_p^2}{\pi\hbar^2(\tau^{-1}-i\omega)},
\label{eq:sigma_intra}
 \ee
 where $\tau$ is determined by impurity scattering and electron-phonon interaction ${\tau ^{ - 1}} = \tau _{imp}^{ - 1} + \tau _{e - ph}^{ - 1}$ .
 Using the mobility $\mu$ of the NPG sheet, it can be presented in the form
 $\tau^{-1}=ev_F^2/(\mu E_F)$, where $v_F=10^6$ m/s is the Fermi velocity in graphene.
 $\omega_p=\sqrt{e^2E_F/2\varepsilon_m}$ is the bulk graphene plasma frequency.
Since the graphene sheets are electronically decoupled from each other by the insulating FeCl$_3$ layers (see above discussion), the optical conductivity of MLG-FeCl$_3$ is given by
\be
\sigma_{\rm intra}^{\rm MLG-FeCl_3}(\omega) = N\frac{e^2}{\pi\hbar^2}\frac{E_F}{\tau^{-1} - i\omega },
\label{eq:sigma_intra^MLG-RuCl3}
 \ee
 where $N$ is the number of graphene layers. This formula is only valid for excitation energies below the band gap $E_g=1$ eV of FeCl$_3$.
 Since the LSP resonances occur at energies between 250 meV and 950 meV, we can safely neglect the optical phonons in graphene at 200 meV and the optical phonons in FeCl$_3$ at 2.7 meV, 7 meV

%The dielectric function for graphene is given by\cite{Safaei2017,Paudel2017}
%\be
%\varepsilon_{||}(\omega)=\varepsilon_g+\frac{i\sigma_{2D}(\omega)}{\varepsilon_0\omega d},
%\label{eq:epsilon}
%\ee
%where $\epsilon_g=2.5$ is the dielectric constant of graphite and $d$ is the thickness of graphene.
%The LSP frequency of the hole can be determined from the equation\cite{Shabbir2020}
%\be
%\varepsilon_m+L_{||}[\varepsilon_{||}(\omega)-\varepsilon_m]=0,
%\label{eq:plasmon}
%\ee
%where $\varepsilon_m$ is the dielectric constant of FeCl$_3$ and $L_{||}$ is a geometric factor.
%Inserting Eq.~(\ref{eq:epsilon}) into this equation gives
%\be
%\varepsilon_m+L_{||}[\varepsilon_g+ i\frac{ne^2}{\pi\hbar^2}\frac{E_F}{\varepsilon_0\omega d(\tau^{-1}-i\omega)}-\varepsilon_m]=0.
%\ee
%Solving for the frequency and using the real part we obtain the LSP frequency,
%\be
%{\rm Re}\omega_{\rm LSP} = \frac{{2nL_{||}^2\varepsilon_m\omega_p^2\tau}}{{\pi\hbar^2 \left\{ {{L_{||}^2} + (2nd)^2\varepsilon_0^2{{\left[ {L_{||}\left( {\varepsilon_g - \varepsilon_m} \right) +\varepsilon_m} \right]}^2}} \right\}}}.
%\ee
%This result shows that the LSP scales linearly with $n$.

\begin{figure*}[htb]
\begin{centering}
\includegraphics[width=18cm]{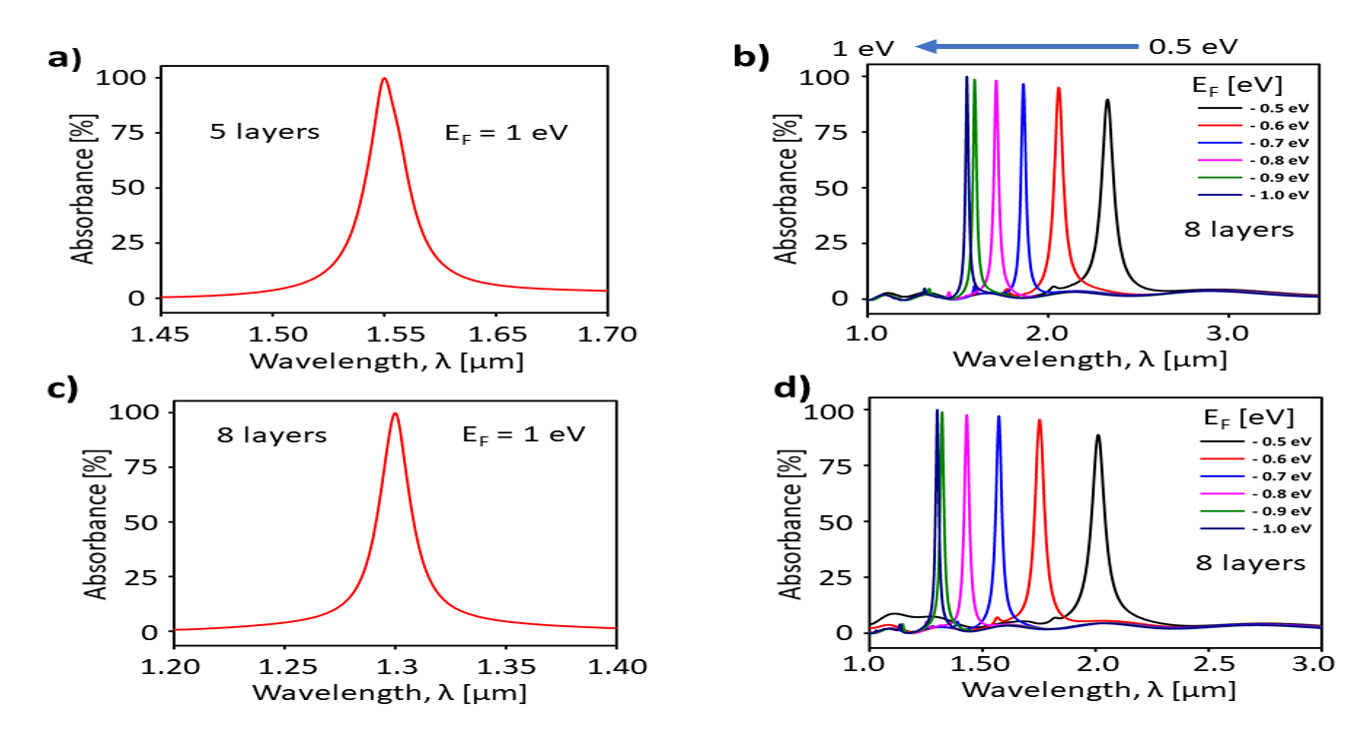}
\end{centering}
\caption{Absorbance as a function of Fermi energy $E_F$. (a) For NPMLG-FeCl$_3$ containing 5  layers of graphene the LSP resonance occurs at $\lambda=1.55$ $\mu$m. (b) The LSP resonance of the 5-graphene layer structure can be tuned from $\lambda=2.5$ $\mu$m to 1.55 $\mu$m by varying the Fermi energy between $E_F=-0.5$ eV and -1.0 eV.
(c) For NPMLG-FeCl$_3$ containing 8  layers of graphene the LSP resonance occurs at $\lambda=1.30$ $\mu$m. (d) The LSP resonance of the 8-graphene layer structure can be tuned from $\lambda=2.0$ $\mu$m to 1.30 $\mu$m  by varying the Fermi energy between $E_F=-0.5$ eV and -1.0 eV.
\label{fig:absorbance_EF} }
\end{figure*}

We use FDTD to calculate the absorbance as a function of Fermi energy $E_F$ for NPMLG-FeCl$_3$ containing $N=5$ and $N=8$ layers of graphene, as shown in Fig.~\ref{fig:absorbance_EF}.
The LSP resonances for the 5-graphene layer and 8-graphene layer structures exhibit wide tunability as a function of the Fermi energy $E_F$.
Compared to our previous results on NPG, the absorbance remains above 80\% even for lower Fermi energies of $E_F=-0.5$ eV.
This large absorbance gives rise to strong heating of the nanopatterned side of the NPMLG structure.

\begin{figure*}[htb]
\begin{centering}
\includegraphics[width=\linewidth]{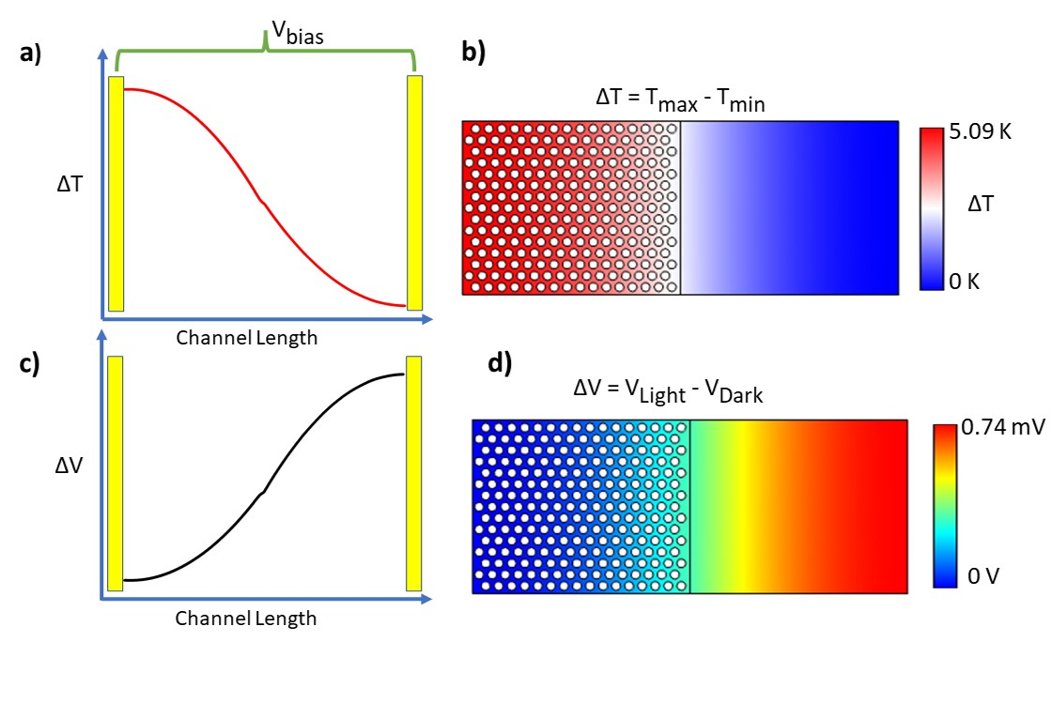}
\end{centering}
\caption{COMSOL simulated temperature gradient and Seebeck voltage generated by the plasmonically enhanced photothermoelectric effect in NPMLG-FeCl$_3$ heterostructure containing $N=5$ graphene layers for the LSP resonance at $\lambda=1.55$ $\mu$m. The channel length and width are both 10 $\mu$m. For an IR light incident power of $P_{\rm inc}=120$ nW a temperature difference of $\Delta T=5.09$ K (a,b) and a Seebeck voltage of $\Delta V=0.74$ mV (c,d) are achieved.
\label{fig:temperature_gradient} }
\end{figure*}

Using COMSOL, we obtain the temperature distribution inside the NPMLG-FeCl$_3$ heterostructure, as shown in Fig.~\ref{fig:temperature_gradient}. 
The photothermoelectric effect in the NPMLG-FeCl$_3$ heterostructure relies on large difference in absorbance between the patterned and the unpatterned side, i.e. the patterned side exhibits a plasmonically enhanced absorbance of nearly 100\% while the unpatterned side absorbs only about 2\% of the incident IR light. 
This results in a temperature gradient across the channel length $\Delta X=X_L-X_R$, where $X_L$ and $X_R$ are the edge positions of the left and right Au contacts, respectively.
The hot carriers created on the patterned side due to Landau damping diffuse to the unpatterned side, resulting in a charge separation and Seebeck voltage
\be
V_S=\int\limits_{X_L}^{X_R} S(x)\frac{\p T_e(x)}{\p x}dx,
\label{eq:Seebeck_voltage}
\ee
where $S(x)$ is the Seebeck coefficient, which has two values $S_{\rm pat}$ and $S_{\rm unpat}$ on the patterned and unpatterned side, respectively.
$T_e(x)$ is the temperature profile of the charge carriers across the channel length, as shown in  Fig.~\ref{fig:temperature_gradient} (b). 
The Seebeck coefficient is approximated well by Mott's formula
\be
S=\frac{\pi^2k_B^2T}{3e}\frac{\p {\rm ln}\sigma}{\p E_F},
\label{eq:Seebeck_coeff}
\ee
where $\sigma$, $k_B$, and $e$ are the electrical conductivity, Boltzmann constant, and elementary charge, respectively.
$\sigma$ and $S$ depend on the $E_F$.

Starting from room temperature at $T=300$ K and an incident power of the IR light of $P_{\rm inc}=120$ nW we obtain a temperature difference of $\Delta T=5.09$ K and a Seebeck voltage of $V_S=0.74$ mV. The Fermi energy of the graphene layers is kept at $E_F=-1.0$ eV, close to the intrinsic p-doping level due to the intercalation with FeCl$_3$.
A bias voltage of $V_b=\pm 0.6$ V is applied for measuring the photocurrents in both directions. Owing to the bias voltage, both plasmonically enhanced photothermoelectric and bolometric effects contribute. In order to remove the bolometric effect for the measurements, the dark and light Seebeck voltages $V_{S,d}$ and $V_{S,l}$ are calculated in the absence and presence of the incident IR light, respectively. The Seebeck voltage is then calculated by $V_S=V_{S,d}-V_{S,l}$.
For a constant Fermi energy $E_F$, a DC bias voltages $+V_b$ and $-V_b$ are applied for two separate simulations across the channel width, which yields the currents $I_>=I+I_S$ and $I_<=-I+I_S$, respectively. $I$ is the current driven by the bias voltage and $I_S=(I_>+I_<)/2$ is the Seebeck current, which is captured in the absence ($I_{S,d}$) and presence ($I_{S,l}$) of the incident IR light. Since the holes are circularly symmetric, $I_S$ is independent of the polarization of the IR light.
The Seebeck current and voltage are then calculated by $I_S=I_{S,l}-I_{S,d}$ and $V_S=RI_S$, respectively, where $R$ is the resistance of the NPMLG-FeCl$_3$ heterostructure.
For the incident IR light, we assume a Gaussian beam with spot size radius of $R_{\rm spot}=2$ mm, a focus size radius of $R_f=\sqrt{(x-x_f)^2+(y-y_f)^2}$, and an incident power of $P_{\rm inc}=120$ nW.
The heat flux through the edges of the holes in the nanopattern is then 
\be
q_0=\frac{2P_{\rm inc}}{\pi R_{\rm spot}^2}e^{-2R_f^2/R_{\rm spot}^2}.
\ee
For an absorbance $A$ obtained from FDTD, the absorbed heat flux is determined by $q_A=Aq_0$.

\begin{figure}[htb]
\begin{centering}
\includegraphics[width=\linewidth]{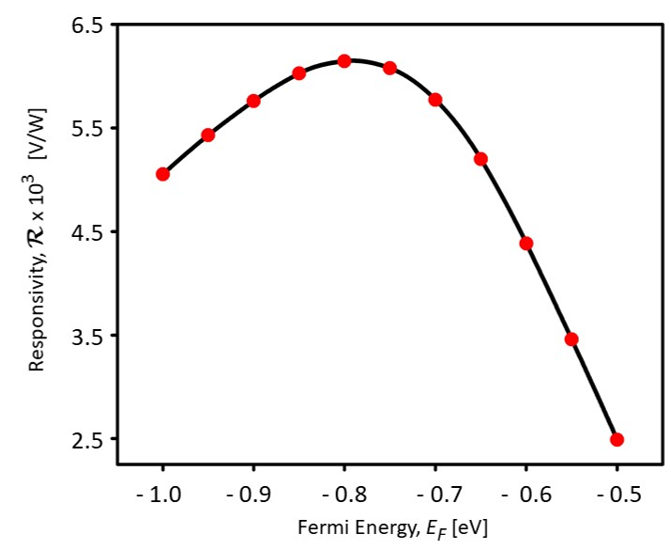}
\end{centering}
\caption{A maximum responsivity of $\calR=6.15\times 10^3$ V/W is achieved at a Fermi energy of $E_F=-0.8$ eV for an incident power of $P_{\rm inc}=120$ nW and a bias voltage of $V_b=0.6$ V.
\label{fig:responsivity} }
\end{figure}

The responsivity is obtained by
\be
\calR=\frac{V_S}{P_{\rm inc}}.
\ee
The responsivity as a function of Fermi energy $E_F$ is shown in Fig.~\ref{fig:responsivity}.
$\calR$ ranges from a minimum of about $2.49\times 10^3$ V/W for $E_F=-0.5$ eV to a maximum of about $6.15\times 10^3$ V/W for $E_F=-0.8$ eV.
Assuming a typical noise equivalent power of ${\rm NEP}=7$ pW/Hz$^{1/2}$ for CVD graphene,\cite{Safaei2019}
we obtain a maximum detectivity of $D^*=\sqrt{LW}/{\rm NEP}=0.74\times 10^9$ Jones.
If we scale the channel width to $W=200$ $\mu$m, it is possible to achieve $D^*=2.33\times 10^9$ Jones.

\section{Conclusion}
In conclusion, we have demonstrated in our theoretical study that NPMLG-FeCl$_3$ can be used to develop a plasmonically enhanced NIR and SWIR photodetector with spectrally tunable selective light absorption. 
Most importantly, the LSPs along with an optical cavity increase the absorbance from about $N\times 2.6$\% for multilayer graphene-FeCl$_3$ (without patterning) to nearly 100\% for NPMLG-FeCl$_3$, where the strong absorbance occurs locally inside the graphene sheets only, thereby outperforming state-of-the-art graphene-based photodetectors.
Remarkably, by taking advantage of NPMLG-FeCl$_3$ with a number of graphene layers $1\leq N\sim 8$ it is possible to develop photodetectors that operate over a wide wavelength range from $\lambda=1.3$ $\mu$m down to $\lambda=12$ $\mu$m and beyond, covering the NIR, SWIR, MWIR, and LWIR regimes.
In the future it would be interesting to study LSPs in NPG also in the THz regime.
Our proposed IR photodetector can be used to develop an IR spectroscopy and detection platform based on NPMLG 
that will be able to detect a variety of molecules that have IR vibrational resonances, such as
CO, CO$_2$, NO, NO$_2$, CH$_4$, TNT, H$_2$O$_2$, acetone, TATP, Sarin, VX, viruses, etc.

\begin{acknowledgments}
We thank Dirk Englund for useful discussions.
\end{acknowledgments}


\begin{thebibliography}{25}%
\makeatletter
\providecommand \@ifxundefined [1]{%
 \@ifx{#1\undefined}
}%
\providecommand \@ifnum [1]{%
 \ifnum #1\expandafter \@firstoftwo
 \else \expandafter \@secondoftwo
 \fi
}%
\providecommand \@ifx [1]{%
 \ifx #1\expandafter \@firstoftwo
 \else \expandafter \@secondoftwo
 \fi
}%
\providecommand \natexlab [1]{#1}%
\providecommand \enquote  [1]{``#1''}%
\providecommand \bibnamefont  [1]{#1}%
\providecommand \bibfnamefont [1]{#1}%
\providecommand \citenamefont [1]{#1}%
\providecommand \href@noop [0]{\@secondoftwo}%
\providecommand \href [0]{\begingroup \@sanitize@url \@href}%
\providecommand \@href[1]{\@@startlink{#1}\@@href}%
\providecommand \@@href[1]{\endgroup#1\@@endlink}%
\providecommand \@sanitize@url [0]{\catcode `\\12\catcode `\$12\catcode
  `\&12\catcode `\#12\catcode `\^12\catcode `\_12\catcode `\%12\relax}%
\providecommand \@@startlink[1]{}%
\providecommand \@@endlink[0]{}%
\providecommand \url  [0]{\begingroup\@sanitize@url \@url }%
\providecommand \@url [1]{\endgroup\@href {#1}{\urlprefix }}%
\providecommand \urlprefix  [0]{URL }%
\providecommand \Eprint [0]{\href }%
\providecommand \doibase [0]{http://dx.doi.org/}%
\providecommand \selectlanguage [0]{\@gobble}%
\providecommand \bibinfo  [0]{\@secondoftwo}%
\providecommand \bibfield  [0]{\@secondoftwo}%
\providecommand \translation [1]{[#1]}%
\providecommand \BibitemOpen [0]{}%
\providecommand \bibitemStop [0]{}%
\providecommand \bibitemNoStop [0]{.\EOS\space}%
\providecommand \EOS [0]{\spacefactor3000\relax}%
\providecommand \BibitemShut  [1]{\csname bibitem#1\endcsname}%
\let\auto@bib@innerbib\@empty
%</preamble>
\bibitem [{\citenamefont {Safaei}\ \emph {et~al.}(2017)\citenamefont {Safaei},
  \citenamefont {Chandra}, \citenamefont {Vázquez-Guardado}, \citenamefont
  {Calderon}, \citenamefont {Franklin}, \citenamefont {Tetard}, \citenamefont
  {Zhai}, \citenamefont {Leuenberger},\ and\ \citenamefont
  {Chanda}}]{Safaei2017}%
  \BibitemOpen
  \bibfield  {author} {\bibinfo {author} {\bibfnamefont {A.}~\bibnamefont
  {Safaei}}, \bibinfo {author} {\bibfnamefont {S.}~\bibnamefont {Chandra}},
  \bibinfo {author} {\bibfnamefont {A.}~\bibnamefont {Vázquez-Guardado}},
  \bibinfo {author} {\bibfnamefont {J.}~\bibnamefont {Calderon}}, \bibinfo
  {author} {\bibfnamefont {D.}~\bibnamefont {Franklin}}, \bibinfo {author}
  {\bibfnamefont {L.}~\bibnamefont {Tetard}}, \bibinfo {author} {\bibfnamefont
  {L.}~\bibnamefont {Zhai}}, \bibinfo {author} {\bibfnamefont {M.~N.}\
  \bibnamefont {Leuenberger}}, \ and\ \bibinfo {author} {\bibfnamefont
  {D.}~\bibnamefont {Chanda}},\ }\href@noop {} {\bibfield  {journal} {\bibinfo
  {journal} {Physical Review B}\ }\textbf {\bibinfo {volume} {96}},\ \bibinfo
  {pages} {165431} (\bibinfo {year} {2017})}\BibitemShut {NoStop}%
\bibitem [{\citenamefont {Safaei}\ \emph
  {et~al.}(2019{\natexlab{a}})\citenamefont {Safaei}, \citenamefont {Chandra},
  \citenamefont {Leuenberger},\ and\ \citenamefont {Chanda}}]{SafaeiACS}%
  \BibitemOpen
  \bibfield  {author} {\bibinfo {author} {\bibfnamefont {A.}~\bibnamefont
  {Safaei}}, \bibinfo {author} {\bibfnamefont {S.}~\bibnamefont {Chandra}},
  \bibinfo {author} {\bibfnamefont {M.~N.}\ \bibnamefont {Leuenberger}}, \ and\
  \bibinfo {author} {\bibfnamefont {D.}~\bibnamefont {Chanda}},\ }\href@noop {}
  {\bibfield  {journal} {\bibinfo  {journal} {Acs Nano}\ }\textbf {\bibinfo
  {volume} {13}},\ \bibinfo {pages} {421} (\bibinfo {year}
  {2019}{\natexlab{a}})}\BibitemShut {NoStop}%
\bibitem [{\citenamefont {Withers}\ \emph {et~al.}(2013)\citenamefont
  {Withers}, \citenamefont {Bointon}, \citenamefont {Craciun},\ and\
  \citenamefont {Russo}}]{Withers2013}%
  \BibitemOpen
  \bibfield  {author} {\bibinfo {author} {\bibfnamefont {F.}~\bibnamefont
  {Withers}}, \bibinfo {author} {\bibfnamefont {T.~H.}\ \bibnamefont
  {Bointon}}, \bibinfo {author} {\bibfnamefont {M.~F.}\ \bibnamefont
  {Craciun}}, \ and\ \bibinfo {author} {\bibfnamefont {S.}~\bibnamefont
  {Russo}},\ }\href@noop {} {\bibfield  {journal} {\bibinfo  {journal} {ACS
  Nano}\ }\textbf {\bibinfo {volume} {7}},\ \bibinfo {pages} {5052} (\bibinfo
  {year} {2013})}\BibitemShut {NoStop}%
\bibitem [{\citenamefont {Bointon}\ \emph {et~al.}(2015)\citenamefont
  {Bointon}, \citenamefont {Jones}, \citenamefont {De~Sanctis}, \citenamefont
  {Hill-Pearce}, \citenamefont {Craciun},\ and\ \citenamefont
  {Russo}}]{Bointon2015}%
  \BibitemOpen
  \bibfield  {author} {\bibinfo {author} {\bibfnamefont {T.~H.}\ \bibnamefont
  {Bointon}}, \bibinfo {author} {\bibfnamefont {G.~F.}\ \bibnamefont {Jones}},
  \bibinfo {author} {\bibfnamefont {A.}~\bibnamefont {De~Sanctis}}, \bibinfo
  {author} {\bibfnamefont {R.}~\bibnamefont {Hill-Pearce}}, \bibinfo {author}
  {\bibfnamefont {M.~F.}\ \bibnamefont {Craciun}}, \ and\ \bibinfo {author}
  {\bibfnamefont {S.}~\bibnamefont {Russo}},\ }\href@noop {} {\bibfield
  {journal} {\bibinfo  {journal} {Scientific Reports}\ }\textbf {\bibinfo
  {volume} {5}},\ \bibinfo {pages} {16464} (\bibinfo {year}
  {2015})}\BibitemShut {NoStop}%
\bibitem [{\citenamefont {Jiang}\ \emph {et~al.}(2017)\citenamefont {Jiang},
  \citenamefont {Kang}, \citenamefont {Cao}, \citenamefont {Xie}, \citenamefont
  {Zhang}, \citenamefont {Chu}, \citenamefont {Liu},\ and\ \citenamefont
  {Banerjee}}]{Jiang2017}%
  \BibitemOpen
  \bibfield  {author} {\bibinfo {author} {\bibfnamefont {J.}~\bibnamefont
  {Jiang}}, \bibinfo {author} {\bibfnamefont {J.}~\bibnamefont {Kang}},
  \bibinfo {author} {\bibfnamefont {W.}~\bibnamefont {Cao}}, \bibinfo {author}
  {\bibfnamefont {X.}~\bibnamefont {Xie}}, \bibinfo {author} {\bibfnamefont
  {H.}~\bibnamefont {Zhang}}, \bibinfo {author} {\bibfnamefont {J.~H.}\
  \bibnamefont {Chu}}, \bibinfo {author} {\bibfnamefont {W.}~\bibnamefont
  {Liu}}, \ and\ \bibinfo {author} {\bibfnamefont {K.}~\bibnamefont
  {Banerjee}},\ }\href@noop {} {\bibfield  {journal} {\bibinfo  {journal} {Nano
  Letters}\ }\textbf {\bibinfo {volume} {17}},\ \bibinfo {pages} {1482}
  (\bibinfo {year} {2017})}\BibitemShut {NoStop}%
\bibitem [{\citenamefont {Zhan}\ \emph {et~al.}(2010)\citenamefont {Zhan},
  \citenamefont {Sun}, \citenamefont {Ni}, \citenamefont {Liu}, \citenamefont
  {Fan}, \citenamefont {Wang}, \citenamefont {Yu}, \citenamefont {Lam},
  \citenamefont {Huang},\ and\ \citenamefont {Shen}}]{Zhan2010}%
  \BibitemOpen
  \bibfield  {author} {\bibinfo {author} {\bibfnamefont {D.}~\bibnamefont
  {Zhan}}, \bibinfo {author} {\bibfnamefont {L.}~\bibnamefont {Sun}}, \bibinfo
  {author} {\bibfnamefont {Z.~H.}\ \bibnamefont {Ni}}, \bibinfo {author}
  {\bibfnamefont {L.}~\bibnamefont {Liu}}, \bibinfo {author} {\bibfnamefont
  {X.~F.}\ \bibnamefont {Fan}}, \bibinfo {author} {\bibfnamefont
  {Y.}~\bibnamefont {Wang}}, \bibinfo {author} {\bibfnamefont {T.}~\bibnamefont
  {Yu}}, \bibinfo {author} {\bibfnamefont {Y.~M.}\ \bibnamefont {Lam}},
  \bibinfo {author} {\bibfnamefont {W.}~\bibnamefont {Huang}}, \ and\ \bibinfo
  {author} {\bibfnamefont {Z.~X.}\ \bibnamefont {Shen}},\ }\href@noop {}
  {\bibfield  {journal} {\bibinfo  {journal} {Advanced Functional Materials}\
  }\textbf {\bibinfo {volume} {20}},\ \bibinfo {pages} {3504} (\bibinfo {year}
  {2010})}\BibitemShut {NoStop}%
\bibitem [{\citenamefont {Yan}\ \emph {et~al.}(2013)\citenamefont {Yan},
  \citenamefont {Low}, \citenamefont {Zhu}, \citenamefont {Wu}, \citenamefont
  {Freitag}, \citenamefont {Li}, \citenamefont {Guinea}, \citenamefont
  {Avouris},\ and\ \citenamefont {Xia}}]{Yan2013}%
  \BibitemOpen
  \bibfield  {author} {\bibinfo {author} {\bibfnamefont {H.}~\bibnamefont
  {Yan}}, \bibinfo {author} {\bibfnamefont {T.}~\bibnamefont {Low}}, \bibinfo
  {author} {\bibfnamefont {W.}~\bibnamefont {Zhu}}, \bibinfo {author}
  {\bibfnamefont {Y.}~\bibnamefont {Wu}}, \bibinfo {author} {\bibfnamefont
  {M.}~\bibnamefont {Freitag}}, \bibinfo {author} {\bibfnamefont
  {X.}~\bibnamefont {Li}}, \bibinfo {author} {\bibfnamefont {F.}~\bibnamefont
  {Guinea}}, \bibinfo {author} {\bibfnamefont {P.}~\bibnamefont {Avouris}}, \
  and\ \bibinfo {author} {\bibfnamefont {F.}~\bibnamefont {Xia}},\ }\href@noop
  {} {\bibfield  {journal} {\bibinfo  {journal} {Nature Photonics}\ }\textbf
  {\bibinfo {volume} {7}},\ \bibinfo {pages} {394} (\bibinfo {year}
  {2013})}\BibitemShut {NoStop}%
\bibitem [{\citenamefont {Shabbir}\ \emph {et~al.}(2021)\citenamefont
  {Shabbir}, \citenamefont {Chandra},\ and\ \citenamefont
  {Leuenberger}}]{Shabbir2021_VO2}%
  \BibitemOpen
  \bibfield  {author} {\bibinfo {author} {\bibfnamefont {M.~W.}\ \bibnamefont
  {Shabbir}}, \bibinfo {author} {\bibfnamefont {S.}~\bibnamefont {Chandra}}, \
  and\ \bibinfo {author} {\bibfnamefont {M.~N.}\ \bibnamefont {Leuenberger}},\
  }\href@noop {} {\bibfield  {journal} {\bibinfo  {journal} {arXiv:}\ ,\
  \bibinfo {pages} {2103.10311}} (\bibinfo {year} {2021})}\BibitemShut
  {NoStop}%
\bibitem [{\citenamefont {Zhang}\ \emph {et~al.}(2009)\citenamefont {Zhang},
  \citenamefont {Tang}, \citenamefont {Girit}, \citenamefont {Hao},
  \citenamefont {Martin}, \citenamefont {Zettl}, \citenamefont {Crommie},
  \citenamefont {Shen},\ and\ \citenamefont {Wang}}]{Zhang2009}%
  \BibitemOpen
  \bibfield  {author} {\bibinfo {author} {\bibfnamefont {Y.}~\bibnamefont
  {Zhang}}, \bibinfo {author} {\bibfnamefont {T.-T.}\ \bibnamefont {Tang}},
  \bibinfo {author} {\bibfnamefont {C.}~\bibnamefont {Girit}}, \bibinfo
  {author} {\bibfnamefont {Z.}~\bibnamefont {Hao}}, \bibinfo {author}
  {\bibfnamefont {M.~C.}\ \bibnamefont {Martin}}, \bibinfo {author}
  {\bibfnamefont {A.}~\bibnamefont {Zettl}}, \bibinfo {author} {\bibfnamefont
  {M.~F.}\ \bibnamefont {Crommie}}, \bibinfo {author} {\bibfnamefont {Y.~R.}\
  \bibnamefont {Shen}}, \ and\ \bibinfo {author} {\bibfnamefont
  {F.}~\bibnamefont {Wang}},\ }\href@noop {} {\bibfield  {journal} {\bibinfo
  {journal} {Nature}\ }\textbf {\bibinfo {volume} {459}},\ \bibinfo {pages}
  {820} (\bibinfo {year} {2009})}\BibitemShut {NoStop}%
\bibitem [{\citenamefont {Cao}\ \emph {et~al.}(2018)\citenamefont {Cao},
  \citenamefont {Fatemi}, \citenamefont {Fang}, \citenamefont {Watanabe},
  \citenamefont {Taniguchi}, \citenamefont {Kaxiras},\ and\ \citenamefont
  {Jarillo-Herrero}}]{Cao2018}%
  \BibitemOpen
  \bibfield  {author} {\bibinfo {author} {\bibfnamefont {Y.}~\bibnamefont
  {Cao}}, \bibinfo {author} {\bibfnamefont {V.}~\bibnamefont {Fatemi}},
  \bibinfo {author} {\bibfnamefont {S.}~\bibnamefont {Fang}}, \bibinfo {author}
  {\bibfnamefont {K.}~\bibnamefont {Watanabe}}, \bibinfo {author}
  {\bibfnamefont {T.}~\bibnamefont {Taniguchi}}, \bibinfo {author}
  {\bibfnamefont {E.}~\bibnamefont {Kaxiras}}, \ and\ \bibinfo {author}
  {\bibfnamefont {P.}~\bibnamefont {Jarillo-Herrero}},\ }\href@noop {}
  {\bibfield  {journal} {\bibinfo  {journal} {Nature}\ }\textbf {\bibinfo
  {volume} {556}},\ \bibinfo {pages} {43} (\bibinfo {year} {2018})}\BibitemShut
  {NoStop}%
\bibitem [{\citenamefont {Wang}\ \emph {et~al.}(2010)\citenamefont {Wang},
  \citenamefont {Ni}, \citenamefont {Liu}, \citenamefont {Liu}, \citenamefont
  {Cong}, \citenamefont {Yu}, \citenamefont {Wang}, \citenamefont {Shen},\ and\
  \citenamefont {Shen}}]{Wang2010}%
  \BibitemOpen
  \bibfield  {author} {\bibinfo {author} {\bibfnamefont {Y.}~\bibnamefont
  {Wang}}, \bibinfo {author} {\bibfnamefont {Z.}~\bibnamefont {Ni}}, \bibinfo
  {author} {\bibfnamefont {L.}~\bibnamefont {Liu}}, \bibinfo {author}
  {\bibfnamefont {Y.}~\bibnamefont {Liu}}, \bibinfo {author} {\bibfnamefont
  {C.}~\bibnamefont {Cong}}, \bibinfo {author} {\bibfnamefont {T.}~\bibnamefont
  {Yu}}, \bibinfo {author} {\bibfnamefont {X.}~\bibnamefont {Wang}}, \bibinfo
  {author} {\bibfnamefont {D.}~\bibnamefont {Shen}}, \ and\ \bibinfo {author}
  {\bibfnamefont {Z.}~\bibnamefont {Shen}},\ }\href@noop {} {\bibfield
  {journal} {\bibinfo  {journal} {ACS Nano}\ }\textbf {\bibinfo {volume} {4}},\
  \bibinfo {pages} {4074} (\bibinfo {year} {2010})}\BibitemShut {NoStop}%
\bibitem [{\citenamefont {Hass}\ \emph {et~al.}(2008)\citenamefont {Hass},
  \citenamefont {Varchon}, \citenamefont {Mill\'an-Otoya}, \citenamefont
  {Sprinkle}, \citenamefont {Sharma}, \citenamefont {de~Heer}, \citenamefont
  {Berger}, \citenamefont {First}, \citenamefont {Magaud},\ and\ \citenamefont
  {Conrad}}]{Hass2008}%
  \BibitemOpen
  \bibfield  {author} {\bibinfo {author} {\bibfnamefont {J.}~\bibnamefont
  {Hass}}, \bibinfo {author} {\bibfnamefont {F.}~\bibnamefont {Varchon}},
  \bibinfo {author} {\bibfnamefont {J.~E.}\ \bibnamefont {Mill\'an-Otoya}},
  \bibinfo {author} {\bibfnamefont {M.}~\bibnamefont {Sprinkle}}, \bibinfo
  {author} {\bibfnamefont {N.}~\bibnamefont {Sharma}}, \bibinfo {author}
  {\bibfnamefont {W.~A.}\ \bibnamefont {de~Heer}}, \bibinfo {author}
  {\bibfnamefont {C.}~\bibnamefont {Berger}}, \bibinfo {author} {\bibfnamefont
  {P.~N.}\ \bibnamefont {First}}, \bibinfo {author} {\bibfnamefont
  {L.}~\bibnamefont {Magaud}}, \ and\ \bibinfo {author} {\bibfnamefont {E.~H.}\
  \bibnamefont {Conrad}},\ }\href@noop {} {\bibfield  {journal} {\bibinfo
  {journal} {Phys. Rev. Lett.}\ }\textbf {\bibinfo {volume} {100}},\ \bibinfo
  {pages} {125504} (\bibinfo {year} {2008})}\BibitemShut {NoStop}%
\bibitem [{\citenamefont {Wang}\ \emph {et~al.}(2020)\citenamefont {Wang},
  \citenamefont {Balgley}, \citenamefont {Gerber}, \citenamefont {Gray},
  \citenamefont {Kumar}, \citenamefont {Lu}, \citenamefont {Yan}, \citenamefont
  {Fereidouni}, \citenamefont {Basnet}, \citenamefont {Yun}, \citenamefont
  {Suri}, \citenamefont {Kitadai}, \citenamefont {Taniguchi}, \citenamefont
  {Watanabe}, \citenamefont {Ling}, \citenamefont {Moodera}, \citenamefont
  {Lee}, \citenamefont {Churchill}, \citenamefont {Hu}, \citenamefont {Yang},
  \citenamefont {Kim}, \citenamefont {Mandrus}, \citenamefont {Henriksen},\
  and\ \citenamefont {Burch}}]{Wang2020}%
  \BibitemOpen
  \bibfield  {author} {\bibinfo {author} {\bibfnamefont {Y.}~\bibnamefont
  {Wang}}, \bibinfo {author} {\bibfnamefont {J.}~\bibnamefont {Balgley}},
  \bibinfo {author} {\bibfnamefont {E.}~\bibnamefont {Gerber}}, \bibinfo
  {author} {\bibfnamefont {M.}~\bibnamefont {Gray}}, \bibinfo {author}
  {\bibfnamefont {N.}~\bibnamefont {Kumar}}, \bibinfo {author} {\bibfnamefont
  {X.}~\bibnamefont {Lu}}, \bibinfo {author} {\bibfnamefont {J.-Q.}\
  \bibnamefont {Yan}}, \bibinfo {author} {\bibfnamefont {A.}~\bibnamefont
  {Fereidouni}}, \bibinfo {author} {\bibfnamefont {R.}~\bibnamefont {Basnet}},
  \bibinfo {author} {\bibfnamefont {S.~J.}\ \bibnamefont {Yun}}, \bibinfo
  {author} {\bibfnamefont {D.}~\bibnamefont {Suri}}, \bibinfo {author}
  {\bibfnamefont {H.}~\bibnamefont {Kitadai}}, \bibinfo {author} {\bibfnamefont
  {T.}~\bibnamefont {Taniguchi}}, \bibinfo {author} {\bibfnamefont
  {K.}~\bibnamefont {Watanabe}}, \bibinfo {author} {\bibfnamefont
  {X.}~\bibnamefont {Ling}}, \bibinfo {author} {\bibfnamefont {J.}~\bibnamefont
  {Moodera}}, \bibinfo {author} {\bibfnamefont {Y.~H.}\ \bibnamefont {Lee}},
  \bibinfo {author} {\bibfnamefont {H.~O.~H.}\ \bibnamefont {Churchill}},
  \bibinfo {author} {\bibfnamefont {J.}~\bibnamefont {Hu}}, \bibinfo {author}
  {\bibfnamefont {L.}~\bibnamefont {Yang}}, \bibinfo {author} {\bibfnamefont
  {E.-A.}\ \bibnamefont {Kim}}, \bibinfo {author} {\bibfnamefont {D.~G.}\
  \bibnamefont {Mandrus}}, \bibinfo {author} {\bibfnamefont {E.~A.}\
  \bibnamefont {Henriksen}}, \ and\ \bibinfo {author} {\bibfnamefont {K.~S.}\
  \bibnamefont {Burch}},\ }\href@noop {} {\bibfield  {journal} {\bibinfo
  {journal} {Nano Letters}\ }\textbf {\bibinfo {volume} {20}},\ \bibinfo
  {pages} {8446} (\bibinfo {year} {2020})}\BibitemShut {NoStop}%
\bibitem [{\citenamefont {Reschke}\ \emph {et~al.}(2017)\citenamefont
  {Reschke}, \citenamefont {Mayr}, \citenamefont {Wang}, \citenamefont {Do},
  \citenamefont {Choi},\ and\ \citenamefont {Loidl}}]{Reschke2017}%
  \BibitemOpen
  \bibfield  {author} {\bibinfo {author} {\bibfnamefont {S.}~\bibnamefont
  {Reschke}}, \bibinfo {author} {\bibfnamefont {F.}~\bibnamefont {Mayr}},
  \bibinfo {author} {\bibfnamefont {Z.}~\bibnamefont {Wang}}, \bibinfo {author}
  {\bibfnamefont {S.-H.}\ \bibnamefont {Do}}, \bibinfo {author} {\bibfnamefont
  {K.-Y.}\ \bibnamefont {Choi}}, \ and\ \bibinfo {author} {\bibfnamefont
  {A.}~\bibnamefont {Loidl}},\ }\href@noop {} {\bibfield  {journal} {\bibinfo
  {journal} {Phys. Rev. B}\ }\textbf {\bibinfo {volume} {96}},\ \bibinfo
  {pages} {165120} (\bibinfo {year} {2017})}\BibitemShut {NoStop}%
\bibitem [{\citenamefont {Warzanowski}\ \emph {et~al.}(2020)\citenamefont
  {Warzanowski}, \citenamefont {Borgwardt}, \citenamefont {Hopfer},
  \citenamefont {Attig}, \citenamefont {Koethe}, \citenamefont {Becker},
  \citenamefont {Tsurkan}, \citenamefont {Loidl}, \citenamefont {Hermanns},
  \citenamefont {van Loosdrecht},\ and\ \citenamefont
  {Gr\"uninger}}]{Warzanowski2020}%
  \BibitemOpen
  \bibfield  {author} {\bibinfo {author} {\bibfnamefont {P.}~\bibnamefont
  {Warzanowski}}, \bibinfo {author} {\bibfnamefont {N.}~\bibnamefont
  {Borgwardt}}, \bibinfo {author} {\bibfnamefont {K.}~\bibnamefont {Hopfer}},
  \bibinfo {author} {\bibfnamefont {J.}~\bibnamefont {Attig}}, \bibinfo
  {author} {\bibfnamefont {T.~C.}\ \bibnamefont {Koethe}}, \bibinfo {author}
  {\bibfnamefont {P.}~\bibnamefont {Becker}}, \bibinfo {author} {\bibfnamefont
  {V.}~\bibnamefont {Tsurkan}}, \bibinfo {author} {\bibfnamefont
  {A.}~\bibnamefont {Loidl}}, \bibinfo {author} {\bibfnamefont
  {M.}~\bibnamefont {Hermanns}}, \bibinfo {author} {\bibfnamefont {P.~H.~M.}\
  \bibnamefont {van Loosdrecht}}, \ and\ \bibinfo {author} {\bibfnamefont
  {M.}~\bibnamefont {Gr\"uninger}},\ }\href@noop {} {\bibfield  {journal}
  {\bibinfo  {journal} {Phys. Rev. Research}\ }\textbf {\bibinfo {volume}
  {2}},\ \bibinfo {pages} {042007} (\bibinfo {year} {2020})}\BibitemShut
  {NoStop}%
\bibitem [{\citenamefont {Sinn}\ \emph {et~al.}(2016)\citenamefont {Sinn},
  \citenamefont {Kim}, \citenamefont {Kim}, \citenamefont {Lee}, \citenamefont
  {Won}, \citenamefont {Oh}, \citenamefont {Han}, \citenamefont {Chang},
  \citenamefont {Hur}, \citenamefont {Sato}, \citenamefont {Park},
  \citenamefont {Kim}, \citenamefont {Kim},\ and\ \citenamefont
  {Noh}}]{Sinn2016}%
  \BibitemOpen
  \bibfield  {author} {\bibinfo {author} {\bibfnamefont {S.}~\bibnamefont
  {Sinn}}, \bibinfo {author} {\bibfnamefont {C.~H.}\ \bibnamefont {Kim}},
  \bibinfo {author} {\bibfnamefont {B.~H.}\ \bibnamefont {Kim}}, \bibinfo
  {author} {\bibfnamefont {K.~D.}\ \bibnamefont {Lee}}, \bibinfo {author}
  {\bibfnamefont {C.~J.}\ \bibnamefont {Won}}, \bibinfo {author} {\bibfnamefont
  {J.~S.}\ \bibnamefont {Oh}}, \bibinfo {author} {\bibfnamefont
  {M.}~\bibnamefont {Han}}, \bibinfo {author} {\bibfnamefont {Y.~J.}\
  \bibnamefont {Chang}}, \bibinfo {author} {\bibfnamefont {N.}~\bibnamefont
  {Hur}}, \bibinfo {author} {\bibfnamefont {H.}~\bibnamefont {Sato}}, \bibinfo
  {author} {\bibfnamefont {B.-G.}\ \bibnamefont {Park}}, \bibinfo {author}
  {\bibfnamefont {C.}~\bibnamefont {Kim}}, \bibinfo {author} {\bibfnamefont
  {H.-D.}\ \bibnamefont {Kim}}, \ and\ \bibinfo {author} {\bibfnamefont
  {T.~W.}\ \bibnamefont {Noh}},\ }\href@noop {} {\bibfield  {journal} {\bibinfo
   {journal} {Scientific Reports}\ }\textbf {\bibinfo {volume} {6}},\ \bibinfo
  {pages} {39544} (\bibinfo {year} {2016})}\BibitemShut {NoStop}%
\bibitem [{\citenamefont {Khrapach}\ \emph {et~al.}(2012)\citenamefont
  {Khrapach}, \citenamefont {Withers}, \citenamefont {Bointon}, \citenamefont
  {Polyushkin}, \citenamefont {Barnes}, \citenamefont {Russo},\ and\
  \citenamefont {Craciun}}]{Khrapach2012}%
  \BibitemOpen
  \bibfield  {author} {\bibinfo {author} {\bibfnamefont {I.}~\bibnamefont
  {Khrapach}}, \bibinfo {author} {\bibfnamefont {F.}~\bibnamefont {Withers}},
  \bibinfo {author} {\bibfnamefont {T.~H.}\ \bibnamefont {Bointon}}, \bibinfo
  {author} {\bibfnamefont {D.~K.}\ \bibnamefont {Polyushkin}}, \bibinfo
  {author} {\bibfnamefont {W.~L.}\ \bibnamefont {Barnes}}, \bibinfo {author}
  {\bibfnamefont {S.}~\bibnamefont {Russo}}, \ and\ \bibinfo {author}
  {\bibfnamefont {M.~F.}\ \bibnamefont {Craciun}},\ }\href@noop {} {\bibfield
  {journal} {\bibinfo  {journal} {Advanced Materials}\ }\textbf {\bibinfo
  {volume} {24}},\ \bibinfo {pages} {2844} (\bibinfo {year}
  {2012})}\BibitemShut {NoStop}%
\bibitem [{\citenamefont {Zhukova}\ \emph {et~al.}(2019)\citenamefont
  {Zhukova}, \citenamefont {Hogan}, \citenamefont {Oparin}, \citenamefont
  {Shaban}, \citenamefont {Grachev}, \citenamefont {Kovalska}, \citenamefont
  {Walsh}, \citenamefont {Cracium}, \citenamefont {Baldycheva},\ and\
  \citenamefont {Tcypkin}}]{Zhukova2019}%
  \BibitemOpen
  \bibfield  {author} {\bibinfo {author} {\bibfnamefont {M.~O.}\ \bibnamefont
  {Zhukova}}, \bibinfo {author} {\bibfnamefont {B.~T.}\ \bibnamefont {Hogan}},
  \bibinfo {author} {\bibfnamefont {E.~N.}\ \bibnamefont {Oparin}}, \bibinfo
  {author} {\bibfnamefont {P.~S.}\ \bibnamefont {Shaban}}, \bibinfo {author}
  {\bibfnamefont {Y.~V.}\ \bibnamefont {Grachev}}, \bibinfo {author}
  {\bibfnamefont {E.}~\bibnamefont {Kovalska}}, \bibinfo {author}
  {\bibfnamefont {K.~K.}\ \bibnamefont {Walsh}}, \bibinfo {author}
  {\bibfnamefont {M.~F.}\ \bibnamefont {Cracium}}, \bibinfo {author}
  {\bibfnamefont {A.}~\bibnamefont {Baldycheva}}, \ and\ \bibinfo {author}
  {\bibfnamefont {A.~N.}\ \bibnamefont {Tcypkin}},\ }\href@noop {} {\bibfield
  {journal} {\bibinfo  {journal} {Nanoscale Research Letters}\ }\textbf
  {\bibinfo {volume} {14}},\ \bibinfo {pages} {225} (\bibinfo {year}
  {2019})}\BibitemShut {NoStop}%
\bibitem [{\citenamefont {Song}\ \emph {et~al.}(2011)\citenamefont {Song},
  \citenamefont {Rudner}, \citenamefont {Marcus},\ and\ \citenamefont
  {Levitov}}]{Song2011}%
  \BibitemOpen
  \bibfield  {author} {\bibinfo {author} {\bibfnamefont {J.~C.~W.}\
  \bibnamefont {Song}}, \bibinfo {author} {\bibfnamefont {M.~S.}\ \bibnamefont
  {Rudner}}, \bibinfo {author} {\bibfnamefont {C.~M.}\ \bibnamefont {Marcus}},
  \ and\ \bibinfo {author} {\bibfnamefont {L.~S.}\ \bibnamefont {Levitov}},\
  }\href@noop {} {\bibfield  {journal} {\bibinfo  {journal} {Nano Letters}\
  }\textbf {\bibinfo {volume} {11}},\ \bibinfo {pages} {4688} (\bibinfo {year}
  {2011})}\BibitemShut {NoStop}%
\bibitem [{\citenamefont {Perdew}\ and\ \citenamefont
  {Zunger}(1981)}]{PZ_functionals}%
  \BibitemOpen
  \bibfield  {author} {\bibinfo {author} {\bibfnamefont {J.~P.}\ \bibnamefont
  {Perdew}}\ and\ \bibinfo {author} {\bibfnamefont {A.}~\bibnamefont
  {Zunger}},\ }\href {\doibase 10.1103/PhysRevB.23.5048} {\bibfield  {journal}
  {\bibinfo  {journal} {Phys. Rev. B}\ }\textbf {\bibinfo {volume} {23}},\
  \bibinfo {pages} {5048} (\bibinfo {year} {1981})}\BibitemShut {NoStop}%
\bibitem [{QW_(2019)}]{QW_1}%
  \BibitemOpen
  \href {http://www.quantumwise.com/} {\bibfield  {journal} {\bibinfo
  {journal} {http://www.quantumwise.com/}\ } (\bibinfo {year}
  {2019})}\BibitemShut {NoStop}%
\bibitem [{\citenamefont {Hashimoto}\ \emph {et~al.}(1989)\citenamefont
  {Hashimoto}, \citenamefont {Forster},\ and\ \citenamefont
  {Moss}}]{Hashimoto1989}%
  \BibitemOpen
  \bibfield  {author} {\bibinfo {author} {\bibfnamefont {S.}~\bibnamefont
  {Hashimoto}}, \bibinfo {author} {\bibfnamefont {K.}~\bibnamefont {Forster}},
  \ and\ \bibinfo {author} {\bibfnamefont {S.~C.}\ \bibnamefont {Moss}},\
  }\href@noop {} {\bibfield  {journal} {\bibinfo  {journal} {Journal of Applied
  Crystallography}\ }\textbf {\bibinfo {volume} {22}},\ \bibinfo {pages} {173}
  (\bibinfo {year} {1989})}\BibitemShut {NoStop}%
\bibitem [{\citenamefont {Li}\ and\ \citenamefont {Yue}(2013)}]{Li2013}%
  \BibitemOpen
  \bibfield  {author} {\bibinfo {author} {\bibfnamefont {Y.}~\bibnamefont
  {Li}}\ and\ \bibinfo {author} {\bibfnamefont {Q.}~\bibnamefont {Yue}},\
  }\href@noop {} {\bibfield  {journal} {\bibinfo  {journal} {Physica B:
  Condensed Matter}\ }\textbf {\bibinfo {volume} {425}},\ \bibinfo {pages} {72}
  (\bibinfo {year} {2013})}\BibitemShut {NoStop}%
\bibitem [{\citenamefont {Paudel}\ \emph {et~al.}(2017)\citenamefont {Paudel},
  \citenamefont {Safaei},\ and\ \citenamefont {Leuenberger}}]{Paudel2017}%
  \BibitemOpen
  \bibfield  {author} {\bibinfo {author} {\bibfnamefont {H.~P.}\ \bibnamefont
  {Paudel}}, \bibinfo {author} {\bibfnamefont {A.}~\bibnamefont {Safaei}}, \
  and\ \bibinfo {author} {\bibfnamefont {M.~N.}\ \bibnamefont {Leuenberger}},\
  }in\ \href@noop {} {\emph {\bibinfo {booktitle} {Nanoplasmonics -
  Fundamentals and Applications}}},\ \bibinfo {editor} {edited by\ \bibinfo
  {editor} {\bibfnamefont {G.}~\bibnamefont {Barbillon}}}\ (\bibinfo
  {publisher} {Intech},\ \bibinfo {address} {London},\ \bibinfo {year} {2017})\
  Chap.~\bibinfo {chapter} {3}, p.\ \bibinfo {pages} {1142}\BibitemShut
  {NoStop}%
\bibitem [{\citenamefont {Safaei}\ \emph
  {et~al.}(2019{\natexlab{b}})\citenamefont {Safaei}, \citenamefont {Chandra},
  \citenamefont {Shabbir}, \citenamefont {Leuenberger},\ and\ \citenamefont
  {Chanda}}]{Safaei2019}%
  \BibitemOpen
  \bibfield  {author} {\bibinfo {author} {\bibfnamefont {A.}~\bibnamefont
  {Safaei}}, \bibinfo {author} {\bibfnamefont {S.}~\bibnamefont {Chandra}},
  \bibinfo {author} {\bibfnamefont {M.~W.}\ \bibnamefont {Shabbir}}, \bibinfo
  {author} {\bibfnamefont {M.~N.}\ \bibnamefont {Leuenberger}}, \ and\ \bibinfo
  {author} {\bibfnamefont {D.}~\bibnamefont {Chanda}},\ }\href@noop {}
  {\bibfield  {journal} {\bibinfo  {journal} {Nature Communications}\ }\textbf
  {\bibinfo {volume} {10}},\ \bibinfo {pages} {3498} (\bibinfo {year}
  {2019}{\natexlab{b}})}\BibitemShut {NoStop}%
\end{thebibliography}
\end{document}